# Shortcut to Adiabatic Passage in a Waveguide Coupler with Allen-Eberly scheme

Koushik Paul and Amarendra K. Sarma*

*Department of Physics, Indian Institute of Technology Guwahati, Guwahati-781039, Assam, India*
*aksarma@iitg.ernet.in

We propose a directional coupler exploiting the framework of adiabatic passage for two level atomic systems with configuration dependent Allen-Eberly scheme. Recently developed shortcut to adiabatic passage method (STA), which uses transitionless quantum driving algorithm, is applied to the coupler. The study shows that it is possible to reduce the device length significantly using STA, keeping the efficiency of power transfer100%.Shortcut approach shows much superiority in terms of robustness and fidelity in power switching compared to the adiabatic one.

## I. INTRODUCTION

Amongst the various popular methods in the fieldof coherent atomic manipulation, techniques that arebased on adiabatic dynamics, such as rapid adiabatic passage (RAP), stimulated Raman adiabatic passage (STIRAP), Stark chirped rapid adiabatic passage (SCRAP)etc. have been studied over a wide range of issues in contemporaryphysics [1-3]. These methods mainly focus on the transferof population among the atomic and molecular states efficiently. Many different processes like controlling chemical reactions, laser cooling, nuclear magnetic resonance(NMR) were realized both theoretically and experimentally in recent past based on adiabatic dynamics [4-6]. Another set of theories in atomic physics, namely transitionless quantum driving[7] and counterdiabatic fieldparadigm[8], has been introduced recently, according towhich adiabatic processes can be driven beyond adiabaticlimit without changing the initial and the final states. These theories enable one todrive a quantum system in infinitely short time without losing robustness of theprocess.It is worthwhile to mention that Chen et al. [9] have proposed a method to speed up the adiabatic passage techniques in the context of two and three-level atoms. This method is now widely termed as shortcuts to adiabatic passage (STA).

Past experiences show that many quantum physics concepts when applied in optics settings can be tested experimentally. To cite some recent examples include parity-time symmetry [10], supersymmetry [11], Anderson localization [12] and so on [13].Recently, based on the analogies between quantum mechanics and wave optics,many techniques have been proposed to manipulate light in various waveguide structures [13-19]. In this regard, waveguide directional couplers in integrated optics are particularly interesting owing to its tremendous practical applications [20,21]. Adiabatic following is applied insuch devices to study the eigenmode evolution of opticalpower through the waveguides [22].For a sufficiently long coupler, where adiabaticity is satisfied, the system follows its initial eigenmode, causing powertransfer from one waveguide to the other. Mode evolution basedstudies of directional couplers show robust optical powerswitching between two, three or even among an arrayof waveguides [22-25]. On the flip side, large device length causes higher transmission loss and makes designing practical devices difficult. However there are significant opportunities to make couplers more efficient and small in dimension using shortcuts to adiabatic passage(STA)[9,26]. Several new studies in this regardhave been reported recently [27-33]. In this work we have studied a directional coupler made of two evanescently coupled waveguides. We propose that these waveguides are designed insuch a way that the waveguide mismatch parameter, Δ, and the coupling parameter, κ, defined later in the article, follows the so calledAllen-Eberly (AE) scheme. It should be noted that the Allen-Eberly scheme for pulse-detuning combination is well established and widely used in atomic physics [9]. For various adiabatic processes AE scheme is much faster compared to other models schemes, such as Landau-Zener scheme [34,35].This article is organized as follows. Sec. II discussesadiabaticity in the coupler, while in Sec.III we discuss how to apply the shortcut technique to the proposed coupler. Sec. IV contains results and discussions followed by conclusions in Sec. V.

## II. ADIABATICITY IN DIRECTIONAL COUPLER

We consider a directional coupler of length $2L$ consisting of two single mode waveguides placed in proximity with propagation constants $\beta_1$ and $\beta_2$. Since we havechosen the guides in close proximity, coupled mode theory can be used to predict the power propagation in thecoupler. In fact it turns out that the prediction ofcoupled mode theory very much resembles the Schrodingerequation for two level atomic system [13]. The coupled equation forthe modal amplitudes $a_1$ and $a_2$of the two waveguides can be written as follows:

$$i \frac{da_1(z)}{dz} = \Delta a_1(z) + \kappa a_2(z) \quad (1)$$
$$i \frac{da_2(z)}{dz} = -\Delta a_2(z) + \kappa a_1(z) \quad (2)$$

Here$\Delta = (\beta_1 - \beta_2)/2$represents the mismatch between the propagation constants and $\kappa$is the coupling between theguides and can be taken to be real without loss of generality.It is easy to see that, in the diabatic basis $\{a_j\}$, there exists an operator similar to the Hamiltonian in quantum physics and can bewritten as:

$$H = \begin{pmatrix} \Delta & \kappa \\ \kappa & -\Delta \end{pmatrix} \quad (3)$$

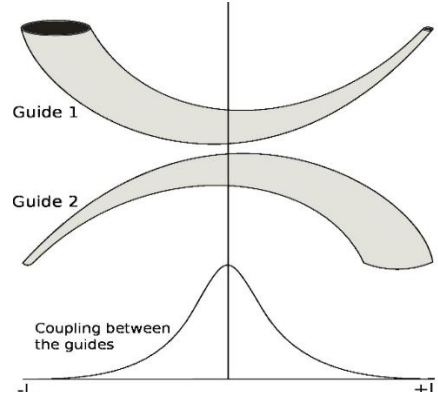

FIG. 1. (Color online) Schematic for adiabatic directional coupler with Allen-Eberly scheme. $\beta_1$ and $\beta_2$are propagation constants for waveguide one and two respectively. Coupler length is $2L$. Maximum of the coupling occurs at $L = 0$.

This Hamiltoniancan be diagonalizedusing unitary transformationto a new basis $\{A_j\}$, which is basically the adiabatic basis, given by

$$\begin{pmatrix} A_1 \\ A_2 \end{pmatrix} = U_0^{-1} \begin{pmatrix} a_1 \\ a_2 \end{pmatrix} (4),$$

where $U_0$ is two-dimensional unitary matrix and can betaken as:

$$U_0 = \begin{pmatrix} \cos(\theta/2) & -\sin(\theta/2) \\ \sin(\theta/2) & \cos(\theta/2) \end{pmatrix} \quad (5)$$

Here $\theta$ is the angle of mixing and is defined as $\tan \theta = \kappa(z)/\Delta(z)$.The transformed Hamiltonian will be:

$$H'(z) = U_0^{-1} H(z) U_0 - i U_0^{-1} \dot{U}_0 \quad (6),$$

where the overdot represents derivative with respect to $z$. Thesecond term is regarded as non-adiabatic correction, owingto fact that the first term is diagonal itself and can drivethe system adiabatically alone. When the adiabatic criterion, which can be written as$\dot{\theta}/2 \ll \sqrt{(\Delta^2 + \kappa^2)}$ ,is satisfied non-adiabatic corrections generally goes to zero.For adiabatic evolution we have followed a coupling-mismatch scheme that is very similar to the famousAllen-Eberly scheme [9] by choosing:

$$\Delta(z) = \Delta_0 \tanh(2\pi z/L) \quad (7)$$
$$\kappa(z) = \kappa_0 \text{sech}(2\pi z/L) \quad (8)$$

The coupling parameter $\kappa(z)$ changes with the coupler length $2L$ and also the mismatch coefficient varies from $-\Delta_0$ to $+\Delta_0$. With both $\Delta$ and $\kappa$ being z dependent, the coupler design results in tapered structure of the waveguides. Moreover the variation of $\Delta(z)$ should be slow enough to accomplish adiabatic evolution. Also for the choices in Eq. (7) and Eq. (8), the adiabatic condition is given by $\kappa_0 L \gg \pi$ and hence it demands the coupler length to be large. The schematic of the proposed coupler is shown in Fig. 1.

## III. REALIZING SHORTCUT

For shortcuts, we followed Berry's algorithm of transitionless quantum driving [7]. Under the circumstances when the adiabatic criterion cannot be fulfilled, complete power switching does not occur due to the effect of the non-adiabatic term in the Hamiltonian. To overcome this we derive a driving Hamiltonian. The Hamiltonian, relevant to our system is simply given by: $H_a = i \sum_j |\partial_z A_j\rangle\langle A_j|$, which when transformed back to the basis $\{a_j\}$, eventually turns out to be

$$H_a = \begin{pmatrix} 0 & -i\dot{\theta}/2 \\ i\dot{\theta}/2 & 0 \end{pmatrix} \quad (9)$$

This additional Hamiltonian should be added back to our original Hamiltonian in order to undo the effects of non-adiabatic terms, which leads to:

$$H_{eff} = \begin{pmatrix} \Delta(z) & \kappa(z) - i\kappa_a(z) \\ \kappa(z) + i\kappa_a(z) & -\Delta(z) \end{pmatrix} \quad (10)$$

This induces an additional coupling, $\kappa_a = \dot{\theta}/2$, with some phase difference with the original one. Also $\kappa_a$ should be comparable with $\kappa$ because the dynamics with $H$ does not need to follow the adiabatic condition. However we can describe it as a combination of an effective coupling and a phase term.

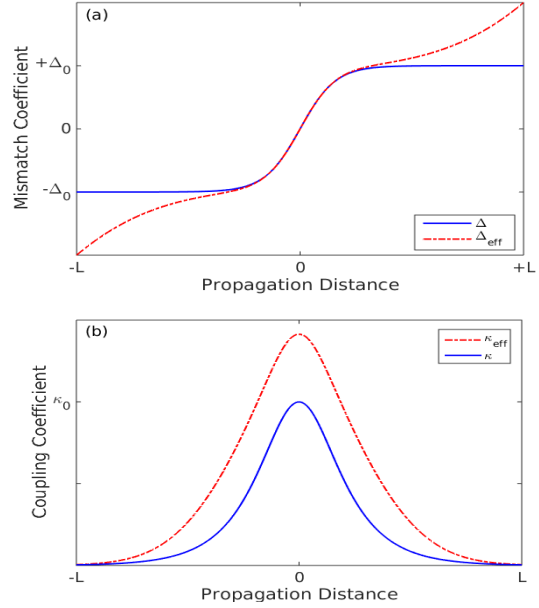

FIG. 2. (Color online) Spatial variation of (a) Mismatch coefficient $\Delta$ and $\Delta_{eff}$. $\Delta$ varies from $-\Delta_0$ to $+\Delta_0$ but $\Delta_{eff}$ is greater towards the ends (b) Coupling coefficients $\kappa$ and $\kappa_{eff}$. Amplitude of $\kappa_{eff}$ is greater than $\kappa$.

$$H_{eff} = \begin{pmatrix} \Delta(z) & \kappa_{eff}(z)e^{-i\phi} \\ \kappa_{eff}(z)e^{i\phi} & -\Delta(z) \end{pmatrix} \quad (11)$$

where $\kappa_{eff} = \sqrt{\kappa^2 + \kappa_a^2}$. Using the following transformation one can eliminate the phase dependence:

$$U_1 = \begin{pmatrix} e^{-i\phi/2} & 0 \\ 0 & e^{i\phi/2} \end{pmatrix} \quad (12),$$

which again provides a new set of basis $\{A_j'\}$ and now the resulting Hamiltonian becomes:

$$H_{eff} = \begin{pmatrix} \Delta_{eff}(z) & \kappa_{eff}(z) \\ \kappa_{eff}(z) & -\Delta_{eff}(z) \end{pmatrix} \quad (13)$$

Here $\Delta_{eff} = \Delta(z) - \dot{\phi}/2$. It is useful to note that $\{A_j'\}$ is related with the adiabatic basis $\{A_j\}$ by two transformations $U_0$ and $U_1$ via the parameters $\theta$ and $\phi$. However, to keep these bases consistent in terms of the initial and the final states, certain conditions need to be imposed. $\theta$ and $\phi$ has to be adjusted in such a way that $\{A_j\}$ and $A_j'$ becomes equivalent at the boundary which leads to the boundary condition $\dot{\theta}(-L) = \dot{\theta}(L) = 0$.

## IV. RESULTS AND DISCUSSIONS

In order to study the power evolution in the coupler we have numerically solved the master equation [1]:

$$\dot{\rho} = -i[H, \rho] \qquad (14)$$

,for both the Hamiltonians in Eq. (3) and Eq. (13). $\rho$ is the density matrix with matrix elements $\rho_{ij} = a_i a_j^*$. Diagonal elements $\rho_{ii} = |a_i(z)|^2$ represents the power in the $i$-th waveguide while the off diagonal elements refers to the coherence between the waveguides. Here the dephasing rate $\Gamma$ has not been considered as we have considered the waveguides to be lossless.

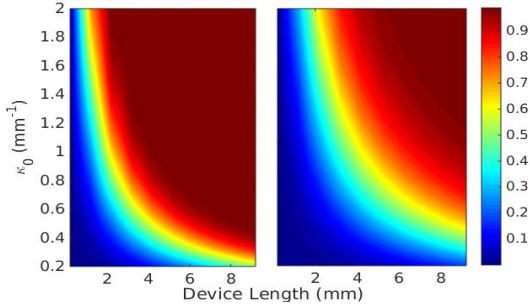

FIG. 3. (Color online)Contour plots for output power with varying $\kappa_0$ and device length $L$. Shortcut (left) shows high fidelity over adiabatic coupler (right).

For adiabatic coupler forms of $\kappa$ and $\Delta$ are taken as in Eq. (7) and Eq. (8). The shortcut approach has been achieved by choosing $\kappa_a$ as follows:

$$\kappa_a = \kappa_0 \exp(-z^2/z_0^2) \qquad (15)$$

Here $z_0$ is the width of the Gaussian. $z_0$ is well adjusted with the varying coupler length so that the boundary conditions for $\theta$ are satisfied at the boundary. Fig. (2a) and Fig. (2b) depicts the spatial variation of the mismatch and coupling parameters for the adiabatic and the STA coupler. The geometry of the coupler depends on the coupling between the waveguides and the mismatch coefficient. Stronger coupling refers to larger separation distances between the waveguides towards the ends of the coupler, which indicates significant decrement in device length. On the other hand the extent of tapering of coupler is controlled by the mismatch coefficient.

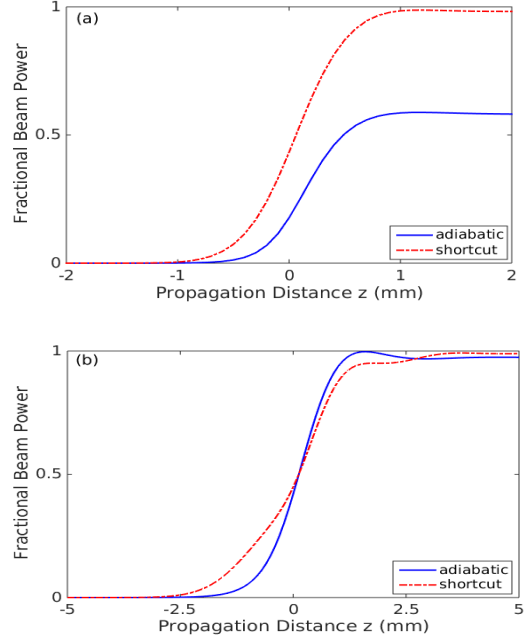

FIG. 4. (Color online)Fractional power output vs. propagation distance for $\Delta_0 = \kappa_0 = 1$ (a) for $2L = 4mm$, (b) for $2L = 10mm$.

As $\Delta$ goes higher, $\beta_1$ and $\beta_2$ tends to change more rapidly throughout the length 2L. As far as adiabaticity is concerned, larger $\kappa(z)$ is preferable for power transfer as it requires to satisfy the condition $\kappa_0 L \gg \pi$. However that does not contribute in shortening of the coupler length. Whereas in STA couplers, significant amount of coupling length can be reduced with little enhanced coupling. These facts can readily be seen in Fig. (3), where we have plotted final power output as a function of device length and the coupling amplitude. With a particular choice of $\Delta_0 = 1\ mm^{-1}$, contours reveals that for large variation of $\kappa_0$ shortcut approach shows much superiority in terms of robustness and fidelity in power switching. For any given value of $\kappa_0$, the minimum distance required to transfer power between the waveguides with adiabatic coupler is much greater, at least two times, than that of the STA coupler. Fig. (4) depicts the spatial evolution of fractional power, defined as $P_2(z)/P_1(-L)$, in the coupler. In our simulation, the input power in the first waveguide is taken to be unity, i.e. $P_1(-L) = 1$, while the input power in second waveguide is kept empty. Other parameters are chosen as: $\Delta_0 = \kappa_0 = 1\ mm^{-1}$.

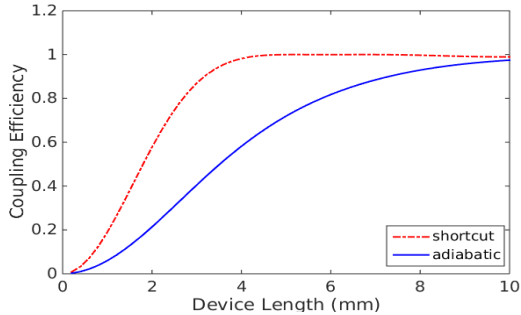

FIG. 5. (Color online)Coupling efficiency for adiabatic and STA coupler with varying device length. Parameters are same as in Fig. (4).

For smaller propagation distance, say $z < 4mm$ or so, the fractional power at the second waveguide, using adiabatic dynamics never reaches unity. It only shows high transfer probability atlarge propagation distances, say $z > 10mm$ or so. However one can achieve nearly 100% power transfer to the second waveguide using the shortcut approach. Coupling efficiency calculation also supports our previous results. Fig (5) illustrates the efficiency of both the adiabatic and the STA coupler with respect to device length. It is quite clear from the plot thatthe STA coupler achieves 100% efficiency with much shorter distance compared to the adiabatic coupler.

## V. CONCLUSION

In conclusion, drawing inspiration from quantum optics, we have proposed a directional coupler based on the Allen-Eberly scheme. The variation in propagation constants $\beta_1(z)$ and $\beta_2(z)$ (and thereby $\Delta$) can be achieved by varying the cross sectional area of the waveguides along the direction of propagation. On the other hand, the coupling parameter, $\kappa$, canbe adjusted by controlling the adjacent distance between the waveguides. The coupler is studied in the adiabatic regime followed by application of recently developed shortcuts to adiabatic passage technique to the coupler. It turns out that by using shortcuts one can reduce the length of the coupler significantly keeping the power transfer efficiency nearly 100%. This study may open new possibilities of exploiting STA and AL scheme for various applications in integrated optics, specifically in the context of photonic circuits.

## ACKNOWLEDGEMENTS

A.K.S. would like to acknowledge the financial support from CSIR (Grant No. 03(1252)/12/EMR-II).